\documentclass[longbibliography,  twocolumn, amsmath, amssymb, aps, prx, superscriptaddress]{revtex4-1}
\usepackage{xr}
\usepackage{amsmath}
\makeatletter
\newcommand*{\addFileDependency}[1]{
  \typeout{(#1)}
  \@addtofilelist{#1}
  \IfFileExists{#1}{}{\typeout{No file #1.}}
}
\makeatother

\newcommand*{\myexternaldocument}[1]{%
    \externaldocument{#1}%
    \addFileDependency{#1.tex}%
    \addFileDependency{#1.aux}%
}
\myexternaldocument{Supplemental}

\usepackage{graphicx}   
\usepackage{dcolumn}  
\usepackage{bm}          
\usepackage{xcolor}    

\usepackage{hyperref}
\hypersetup{
    colorlinks=true,
    linkcolor=blue,
    filecolor=black,      
    urlcolor=blue,
    citecolor=blue,
}
\usepackage{natbib}
\usepackage{braket}
\usepackage{soul}
\usepackage{relsize}
            
\usepackage{pifont}

\begin{document}


\title{Non-monotonic Rheology and Stress Heterogeneity in Confined Granular Suspensions}
\author{Haitao Hu}\affiliation{Department of Physics, The Hong Kong University of Science and Technology, Hong Kong, China}
\author{Yiqiu Zhao}\affiliation{Department of Physics, The Hong Kong University of Science and Technology, Hong Kong, China}
\author{Weiwei Zhao}\affiliation{Department of Physics, The Hong Kong University of Science and Technology, Hong Kong, China}
\author{Ligen Qiao}\affiliation{Lushan New Materials Co.,~Ltd, Guangzhou, China}

\author{Qin Xu}
\email[]{qinxu@ust.hk}
\affiliation{Department of Physics, The Hong Kong University of Science and Technology, Hong Kong, China}

\begin{abstract}

We systematically investigated the impact of boundary confinement on the shear-thickening rheology of dense granular suspensions. Under highly confined conditions, dense suspensions were found to exhibit size-dependent or even rarely reported non-monotonic ($S$-shaped) flow curves in steady states. By performing {\em in-situ} boundary stress microscopy measurements, we observed enhanced flow heterogeneities in confined suspensions, where concentrated high-stress domains propagated stably either along or against the shear direction. By comparing the boundary stress microscopy results with macroscopic flow responses, we revealed the connection between non-monotonic rheology and stress heterogeneity in confined suspensions. These findings suggest the possibility of controlling suspension rheology by imposing different boundary confinements. 

\end{abstract}
\date{\today}
\maketitle

\section{Introduction}

Dense granular suspensions comprise concentrated non-Brownian particles mixed with a Newtonian liquid. They are omnipresent in diverse natural phenomena and engineering applications, including mudflows in landslides~\cite{DeBlasio2009}, construction materials~\cite{Emanuela2022}, and field-responsive fluids~\cite{Vicente2011,Tam1997}. Unlike Newtonian fluids, the rheological response of a dense granular suspension to shear is often nonlinear. When suspension viscosity increases under shear, the  flow behavior is referred to as shear thickening~\cite{Morris2020, Brown2014, Christopher2022}.

The physical mechanism of shear thickening can be understood across a hierarchy of length scales. Recent studies have interpreted shear thickening in dense non-Brownian suspensions has been interpreted as a shear-induced crossover from an unconstrained flow state to a constrained frictional state~\cite{Wyart2014,Mari2015, Guy2015}. Within the shear-thickened flows, the suspended particles form direct contacts by overriding a critical stress, depending on the intrinsic interactions among particles~\cite{Royer2016,Park2019}. The continuous strengthening of large-scale contact networks results in a sharp increase in flow resistance~\cite{Romain2019, Pradeep2021}. This phenomenological framework not only captures the essential characteristics of shear thickening in dense suspensions~\cite{Royer2016}  but also predicts the emergence of $S$-shaped flow curves above a critical volume fraction~\cite{Wyart2014, Romain2015}. However, these non-monotonic rheological responses have  rarely been observed in steady state measurements~\cite{Pan2015}. Instead, unstable flow instabilities and substantial stress fluctuations often dominates the flow responses in this regime~\cite{Xu2020, Rijan2021, Lootens2003, Sedes2020}.

Besides the multi-scale contact networks within dense suspensions, shear boundaries play essential roles in shear thickening. As dense suspensions tend to dilate under shear~\cite{Brown2012, Xu2014}, the shear plates apply positive normal stresses to stablize the contact networks, enabling sustained thickening responses~\cite{Royer2016,seto2018}. Further, the geometric profiles of shear boundaries has been shown to control statistical features, including stress fluctuations~\cite{Xu2020} and flow heterogeneities~\cite{Guillaume2020, Brice2018, Fall2010}, in dense suspensions. However, despite these experimental findings, the underlying mechanism by which the boundary conditions modulate the shear thickening flows remains unclear.

In this work, we systematically characterized the role of boundary confinement in both {\em sedimenting} and {\em non-sedimenting} granular suspensions. By combining rheological characterizations with boundary stress microscopy, we show that a strong confinement imposed by shear boundaries effectively stablizes non-monotonic flow curves and  induce local stress concentrations. Driven by the interplay between shear boundaries and particle interactions, the high-stress domains propagate dynamically, either along or against the shear direction. These emergent features of confined suspensions are greatly reduced when not confined.

\section{Materials}

Athermal granular particles are commonly subjected to the influence of gravity. Therefore, particle-sedimentation plays an essential role in the rheological behaviors of many granular suspensions~\cite{Brown2012, ovarlez2006,Xu2020,Hermes2016}. On the other hand, for particles having polymeric components,  it is possible to match their densities to that of the solvent~\cite{Han2016,Fall2010,Royer2016,Park2019}.  Non-sedimenting suspensions are usually stabilized by intrinsic inter-particle repulsions, such as electrostatic~\cite{Kaldasch2008} and steric interactions~\cite{Kaldasch2009}. In this study, we investigated the shear thickening rheology of both systems under confinement.

The non-sedimenting suspensions were prepared using polystyrene (PS) particles having a diameter of approximately $d_{ps}= 25$~$\mu$m, suspended in a glycerol-water mixture. The solvent made of 20~wt\%  glycerol dissolved in 80~wt\% deionized water had a density of $\rho_{gw} = 1.05$~g/cm$^3$, precisely matching that of PS particles, and a viscosity of approximately 1.75~cSt. The PS particles can be well dispersed and stabilized by electrostatic repulsions in the suspensions~\cite{rathee2021}. To demonstrate this, we added sodium chloride (NaCl) to PS-glycerol suspensions.  For a constant volume fraction of particles ($\phi=60~\%$), the yield stress rose as the salt concentration increased from 0.10~mol/L to 0.63~mol/L (Supplementary Fig.~S1), confirming the role of charge stabilization in PS-glycerol suspensions .

The  sedimenting suspensions were prepared by mixing glass beads ($d_g=43$~$\mu$m) with 20~cSt silicone oil. Due to the density difference between glass ($2.33$~g/cm$^3$) and silicone ($0.95$~g/cm$^3$), the particles were consistently affected by gravity. Thus, the glass particles sedimented in static states, leaving a thin fluid layer between the particles and the shear plate. As the shear stress increased, the particles gradually lifted against gravity~\cite{Xu2020, Brown2012}. In this system, shear thickening occurred as the flow of particles spans across the whole shear gap~\cite{Xu_2014}.

\begin{figure}[t]
	\centering
	\includegraphics[width =87 mm]{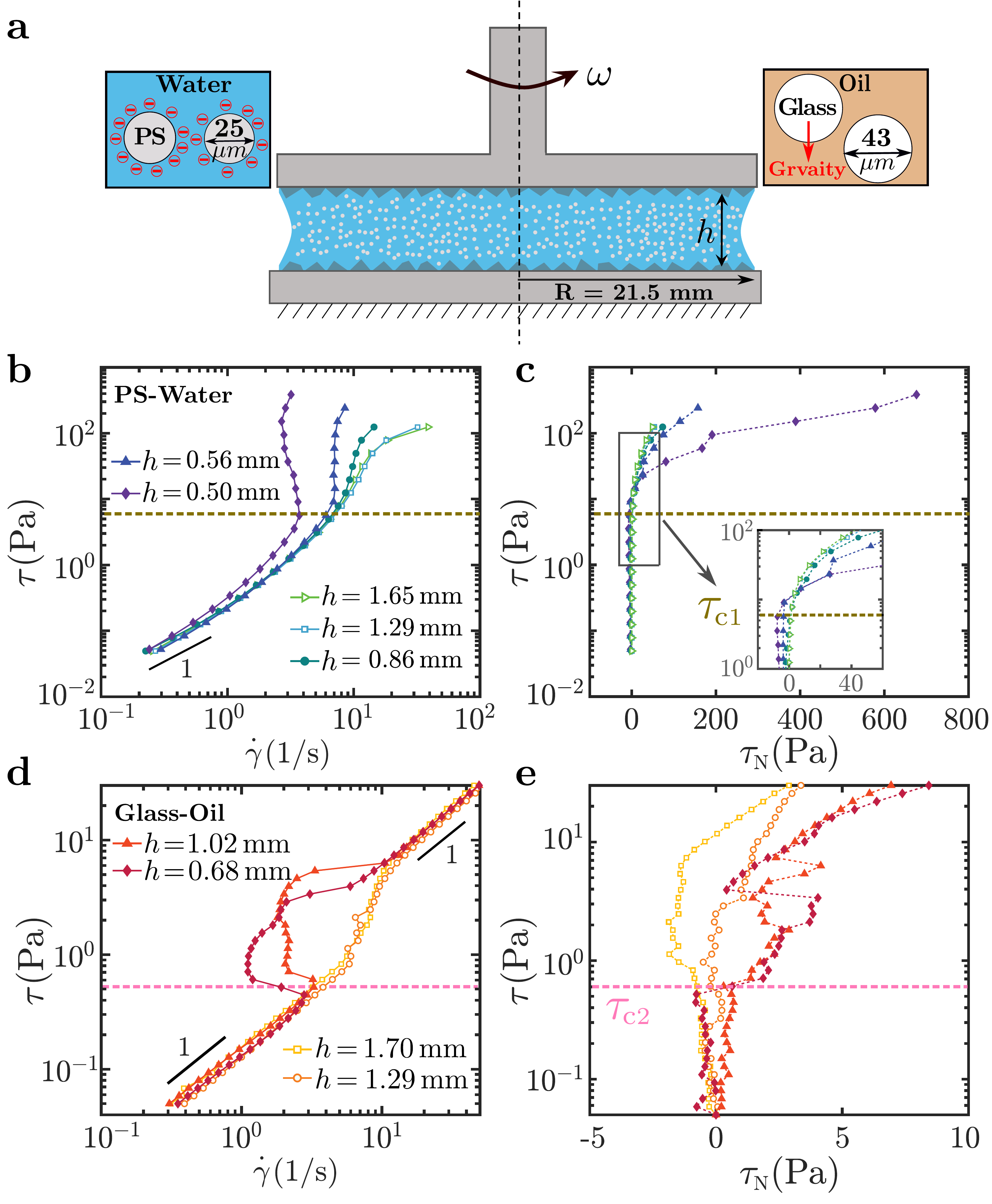}	
	\caption{{\bf Flow curves of confined dense suspensions.} {\bf (a)} Schematic illustration of the parallel-plate shear cell used to measure the shear rheology of dense granular suspensions. {\bf (b)} Plots of shear stress ($\tau$) against the shear rate ($\dot\gamma$) for the non-sedimenting PS-water suspensions at different gap sizes, namely $h = $ 1.65~mm, 1.29~mm, 0.86~mm, 0.56~mm, and 0.50~mm. {\bf (c)} Plots of shear stress ($\tau$) against the corresponding normal stress ($\tau_{\rm N}$) for the PS-water suspensions at the same gap sizes as in panel (b). The inset shows the same data near the onset of shear thickening ($\tau_{c1}$) using expanded axes for clarity. {\bf (d)} Plots of shear stress ($\tau$) against shear rate ($\dot\gamma$) for the sedimenting glass-oil suspensions at different gap sizes, including $h = $ 1.70~mm, 1.29~mm, 1.02~mm, and 0.68~mm. {\bf (e)} Plots of shear stress ($\tau$) against the normal stress ($\tau_{\rm N}$) for the glass-oil suspensions at the same gap sizes shown in panel (d). The brown (panels {\bf (b)} and {\bf (c)}) and pink (panels {\bf (d)} and {\bf(e)}) dashed lines indicate the critical stresses ($\tau_{c1}$ and $\tau_{c2}$) at which the $S$-shapes begin to emerge for  the PS-water suspensions and glass-oil suspensions, respectively.}
	\label{fig:S_curves}
\end{figure}  

\section{Shear rheology of confined suspensions}

{\em Steady-state rheological measurements.---} The shear rheology of both suspensions was characterized using an Anton Paar rheometer (MCR 302) equipped with a parallel-plate shear cell having a diameter of 43~mm (Fig.~\ref{fig:S_curves}(a)). To prevent wall slippage, both shear plates were roughened using sandpapers (P-120 grade) with an average grain size of approximately 120~$\mu$m. We will show that these boundary conditions are essential to obtaining reliable rheological results~\cite{Brice2018, Fall2010}. The shear stress ($\tau$)  was obtained from the torque $T$ measured using the rheometer, according to 
\begin{equation}
\tau= {4 T}/{3 \pi R^{3}}, 
\label{eqn:stress}
\end{equation}
where $R$ is the radius of the shear plate.
The shear rate was calculated from the angular velocity $\omega$ using 
\begin{equation}
\dot\gamma = 2\omega R/(3h), 
\label{eqn:rate}
\end{equation}
where $h$ is the gap size. The viscosity of suspensions was defined by the ratio of shear stress to shear rate, 
\begin{equation}
\eta=\tau / \dot{\gamma}={2 T h} / {\omega \pi R^{4}}. 
\label{eqn:viscosity}
\end{equation}
We simultaneously measured the normal force ($F_N$) applied to the shear plate, resulting in a normal stress 
\begin{equation}
\tau_N = F_N/\pi R^2. 
\end{equation}
A positive normal stress ($\tau_N >0$) indicates that the suspension pushes against the shear plate, whereas a negative normal stress ($\tau_N<0$) indicates the opposite.

{\em PS-water suspensions.---} Figure~\ref{fig:S_curves}(b) shows the flow curves of PS-water suspensions with $\phi=0.60$, where the gap size was varied from $h = 1.65$~mm to $0.50$~mm. For $h = 1.65$~mm (open left-triangles) and $h=1.29$~mm (open circles), the suspension exhibited a Newtonian-like flow behavior, $\tau \sim \dot\gamma$, below a critical stress $\tau_{c1} \approx 6$~Pa. For $\tau>\tau_{c1}$, $\tau(\dot\gamma)$ displayed classical shear thickening responses, where $\tau_{c1}$ is the  stress required to establish direct contacts among PS particles by overcoming the electrostatic repulsion.  

By further decreasing the gap size, we obtained size-dependent flow curves. For instance, the suspensions shear-thicken more profoundly under $h = 0.86$~mm (solid circles), and even discontinuously at $h=0.56$~mm (solid up-triangles). More strikingly, a non-monotonic flow curve of $\tau(\dot\gamma)$ appeared as $h$ became as small as 0.50~mm (solid diamonds), where a flow regime with a negative slope, ${d\tau}/{d\dot\gamma}<0$, emerges above $\tau_c$.

In addition, the increase in the viscosity of PS-water suspensions was associated with a synchronous rise of the normal stress ($\tau_N$). As shown in  Fig.~\ref{fig:S_curves}(c), $\tau_N$ appears positive and increases with $\tau$ above $\tau_{c1} \approx 6$~Pa in all the measurements. For the suspensions with $h=0.56$~mm and 0.50~mm, containing approximately $20$ layers of PS particles, the increase in $\tau_N$ due to thickening-induced dilation became appreciable. Given that $\tau_N$ scales almost linearly with $\tau$ under highly confined conditions (Supplementary Fig.~S2), we conjecture that the discontinuous and non-monotonic flow curves in Fig.~\ref{fig:S_curves}(b) emerge as the system size approaches the correlation length scale of frictional contact networks in the sheared PS-water suspensions.

{\em Glass-oil suspensions.---} We further characterized the rheology of sedimenting glass-oil suspensions as the gap size was varied from $h=1.70$~mm to 0.68~mm. Due to the slow relaxation dynamics of the glass particles in viscous silicone oil, the shear rate $\dot\gamma$ at each shear stress $\tau$ was obtained by averaging the instantaneous values over a period of $t_{\rm w}=200$~s. As a further increase in $t_{\rm w}$ did not vary the flow curves (Supplementary Fig.~S3), we conclude that the resulting $\tau(\dot\gamma)$ represented the flow curves at steady states. Figure~\ref{fig:S_curves}(d) shows $\tau(\dot\gamma)$ of the glass-oil suspensions with $\phi=52$~\% measured at different $h$. For each individual trace, we identified two distinctive Newtonian regimes at $\tau<\tau_{c2} =0.5$~Pa and $\tau > \tau_e = 6$~Pa, respectively. The crossover between the two flow regimes represents a shear thickening transition from a sedimenting to a suspended flow state. As $\tau_{c2}$ is consistent with the gravitational stress $\Delta  \rho g d_g \approx 0.58$~Pa where $\Delta  \rho$ is the density difference between the glass and silicone oil, shear thickening in glass-oil suspensions was initiated by rearranging the particle configurations against sedimentation~\cite{Xu2020, Brown2012}. Similar to the PS-water suspensions, {\em S}-shaped flow curves for glass-oil suspensions emerged as the gap size was reduced to $\sim20$ particle diameters: for $h = 1.02$~mm and $0.68$~mm, $\tau(\dot\gamma)$ becomes non-monotonic between $\tau_{c2}$ and $\tau_e$, with $\tau_N$ rising more drastically than for large gaps (Fig.~\ref{fig:S_curves}(e)).

\begin{figure}[t]
	\centering
	\includegraphics[width =80mm]{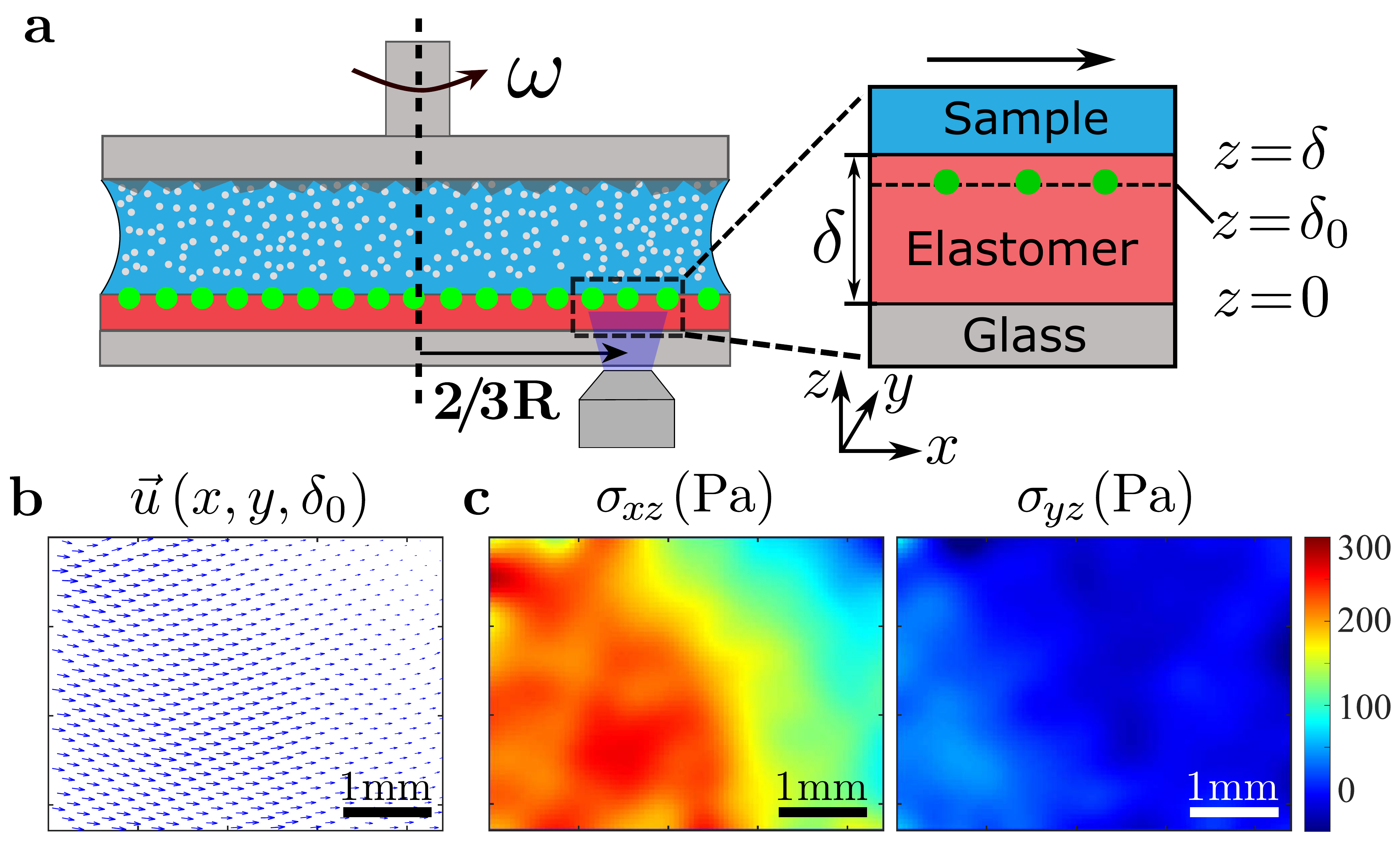}
	\caption{{\bf Boundary stress microscopy (BSM).} (a) Schematic illustration of the boundary stress microscopy incorporated into the parallel-plate shear cell. A film of soft silicone gel coated with fluorescent nanobeads was attached on the bottom shear plate. (b) Representative displacement fields $\vec{u}(x,y)$, and stress maps $\sigma_{xz}$ and $\sigma_{yz}$ measured by BSM.}	
	\label{fig:BSM}
\end{figure}

\begin{figure*}[t]
	\centering
	\includegraphics[width = 176 mm]{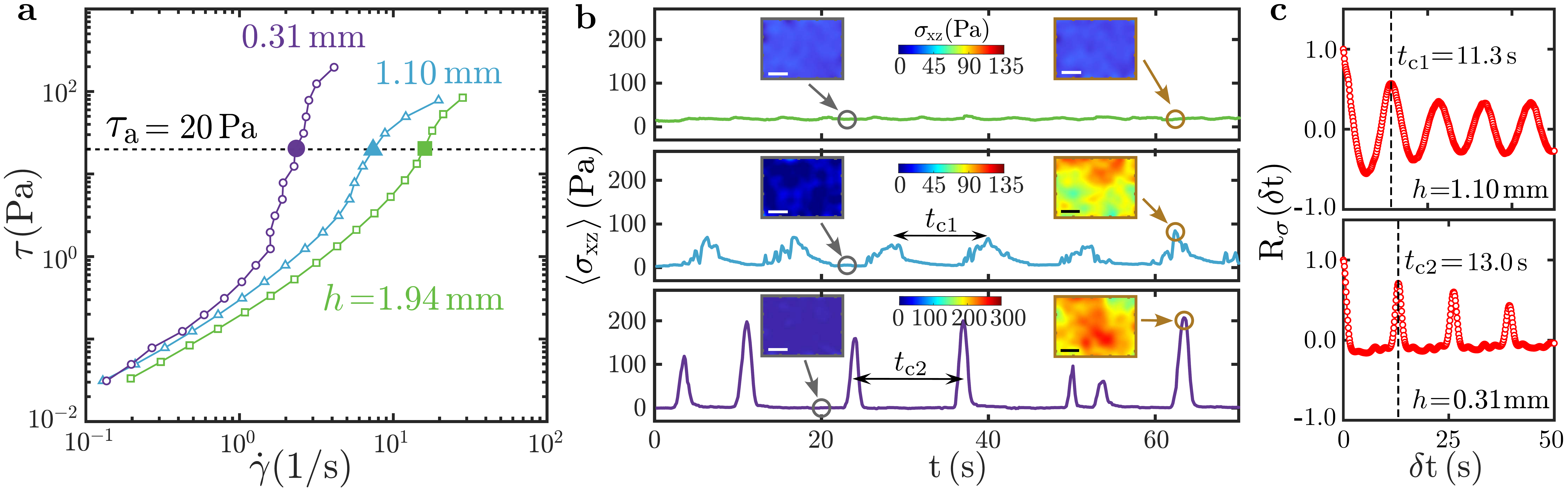}
     \caption{{\bf Local stress fluctuations in PS-water suspensions}. {\bf (a)} Flow curves of the PS-water suspensions measured with the BSM setup at $h =$ 1.94~mm, 1.10~mm, and 0.31~mm. {\bf (b)} Temporal fluctuations of $\langle \sigma_{xz} (t) \rangle$ measured at $\tau_a = 20$~Pa for the gap sizes  $h = $ 1.94~mm (green), 1.10~mm (blue), and 0.31~mm (purple). The characteristic timescales $t_{c1}$ and $t_{c2}$ represent the time intervals between two high-stress events. The insets show the stress map $(\sigma_{xz})$ at the time points indicated by the open circles. Scale bars: $1000$~$\mu$m. {\bf (c)} Autocorrelation function  $R_\sigma(\delta t)$ for $h =$ 1.10~mm and 0.31~mm.} 
	\label{fig:PS oscillations} 
\end{figure*}

{\em Stability of the {\em S}-shaped flow curves.---} As the shear stress $\tau$ was ramped up and down cylically, we observed negligible hysteresis of $\tau(\dot\gamma)$ in the PS-water suspensions (Supplementary Fig.~S4(a)). In contrast,  for glass-oil suspensions under cyclic testing, the {\em S}-shaped traces still remained qualitatively unchanged although the hysteresis loops emerged (Supplementary Fig.~S4(b)). These observations suggest that the non-monotonic flow curves, which have been rarely observed in experiments~\cite{Pan2015}, were stabilized by the imposed confinements. This mechanical stability is consistent with predictions in a previous simulation~\cite{Mari2015_PRE}: a confined suspension can potentially enhance particle contacts while reducing the shear rate, in response to an increasing  shear stress.

We further examined the role of boundary roughness in the non-monotonic responses. Using a shear plate roughened by a finer sand paper (grain size $\approx 50~\mu$m), the $S$-shaped flow curves of PS-water suspensions disappeared at $h = 0.50$~mm and then re-emerged as $h$ was further reduced to $0.33$~mm (Supplementary Fig.~S5(a)). This observation suggests the essential roles of boundary roughness in stabilizing non-monotonic flows in the PS-water suspensions. With a reduced boundary roughness, wall slips can destabilize non-monotonic flows~\cite{Han2019}. In contrast, the {\em S}-shaped flow curves in confined glass-oil suspensions have a negligible dependence on the boundary roughness (Supplementary Fig.~S5(b)). As the glass particles are constantly affected by gravity, the shear plate is more likely to be in contact with the solvent film than with the particles, such that  wall slips are prevented regardless of the boundary roughness.

\section{Local shear stress fluctuations}

{\em Boundary stress microscopy.---} To search for the underlying origin of the non-monotonic behaviors, we employed boundary stress microscopy (BSM) to characterize the shear-induced heterogeneity in confined suspensions. This technique was initially developed to measure in-plane tractions at soft interfaces~ \cite{Xu2010, Mertz2012}, and was later improved for {\em in-situ} measurements of boundary stresses in rheological characterizations~\cite{Rathee2017, Rathee2020}. Below, we briefly summarize the theoretical and experimental methods used in the BSM analysis.

To perform BSM, we coated a layer of soft gels with a Young's modulus $E$ and a thickness $\delta_0$ on the bottom plate of the shear cell. We characterized the local shear stresses at the boundaries of suspensions by measuring the deformations of this soft substrate. Considering linear elasticity, the constitutive relation between stress $ \boldsymbol{\sigma} $ and displacement $\boldsymbol{u}$ of soft gels is expressed as
\begin{equation}
  \boldsymbol{\sigma} = \frac{E}{1+\nu} \left(\frac{1}{2}(\nabla \boldsymbol{u}+\nabla \boldsymbol{u}^T)+\frac{\nu \nabla \cdot \boldsymbol{u}}{1-2 \nu} \boldsymbol{I} \right),
\label{eq:consitutive_relation}  
\end{equation}
where $\nu$ is Poisson's ratio of the soft layer and  $\boldsymbol{I}$ is the unity matrix. In addition, the mechanical equilibrium within the soft layer is governed by
\begin{equation}
	(1-2\nu) \nabla^2  \boldsymbol{u} + \nabla(\nabla\cdot\boldsymbol{u} )=0
\label{eq:mechanical_equilibrium}
\end{equation}
with the boundary conditions
\begin{equation}
\boldsymbol{u}\vert_{z=\delta}=(u_x^*, u_y^*,0) \text { and } \boldsymbol{u}\vert_{z=0}=(0, 0, 0),
\label{eq:boundary_condition}
\end{equation}
where $(u_x^*, u_y^*)$ represents the in-plane displacements on the upper surface of the soft layer~$(z=h)$, whereas the bottom surface is always undeformed~$(z=0)$. As depicted in Fig.~\ref{fig:BSM}(a),  $x$ and $y$ are the shear and vorticity directions, respectively, and  $z$ is the gradient direction. By applying Fourier transformations to both the stress and displacement fields, $\sigma_{ij} = \int \hat \sigma_{ij} \exp[i(k_x x +k_y y)] dk_x dk_y$ and $u_{i} = \int \hat u_{i} \exp[i(k_x x +k_y y)] dk_x dk_y$ where $i, j = x, y, z$, we obtain the relationship between stress and displacement in Fourier space
\begin{equation}
\hat\sigma_{iz}(k_x, k_y, z) = Q_{ij}(k_x, k_y, z) \hat{u}_j (k_x, k_y, z),~(i=x,y),
\label{eq:tensor_product}
\end{equation}
where the explicit form of the matrix $\boldsymbol{Q}$ is provided in the Supplementary Information. As the surface displacements $u_x^*$ and $u_y^*$ are measured experimentally, we quantitatively determine the elastic stresses ($\sigma_{xz}$ and $\sigma_{yz}$) at $z=\delta_0$ using Eq.~\ref{eq:tensor_product} (Fig.~\ref{fig:BSM}(a)). For each stress map obtained by BSM, we denote $\langle \sigma_{iz} \rangle$ as the spatially averaged $\sigma_{iz}$ over the field of view.

\begin{figure*}[t]
	\centering
	\includegraphics[width = 160 mm]{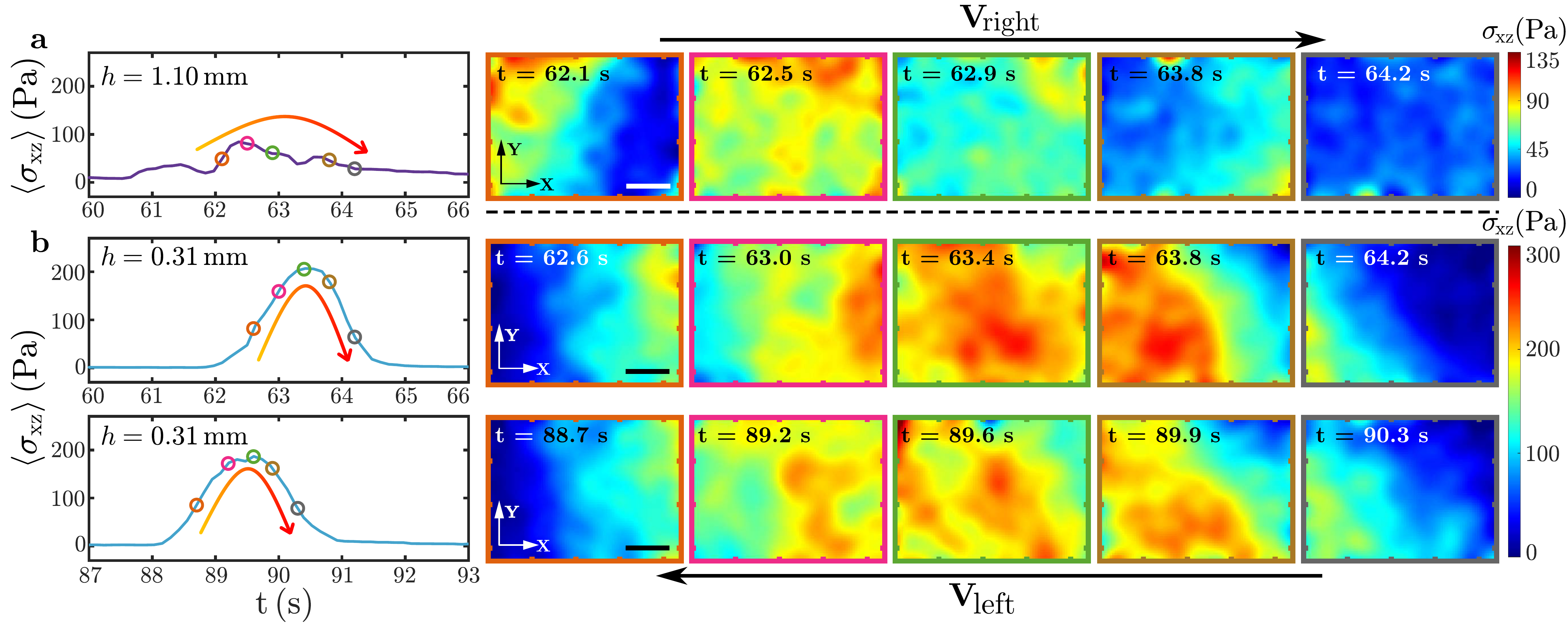}	
	\caption{{\bf Propagation of  high-stress regions in PS-water suspensions for {\bf (a)} $h = $ 1.10~mm and {\bf (b)} $h = $ 0.31~mm}. Left panels: representative peaks of $\langle \sigma_{xz} (t) \rangle$ at $\tau_a = 20$~Pa. Right panel: stress maps of $\sigma_{xz}$ at different times indicated by the open circles in the left panels. Scale bars: $1000$~$\mu$m.} 
	\label{fig:PS front}
\end{figure*}

\begin{figure}[h]
 	\centering
	\includegraphics[width = 70mm]{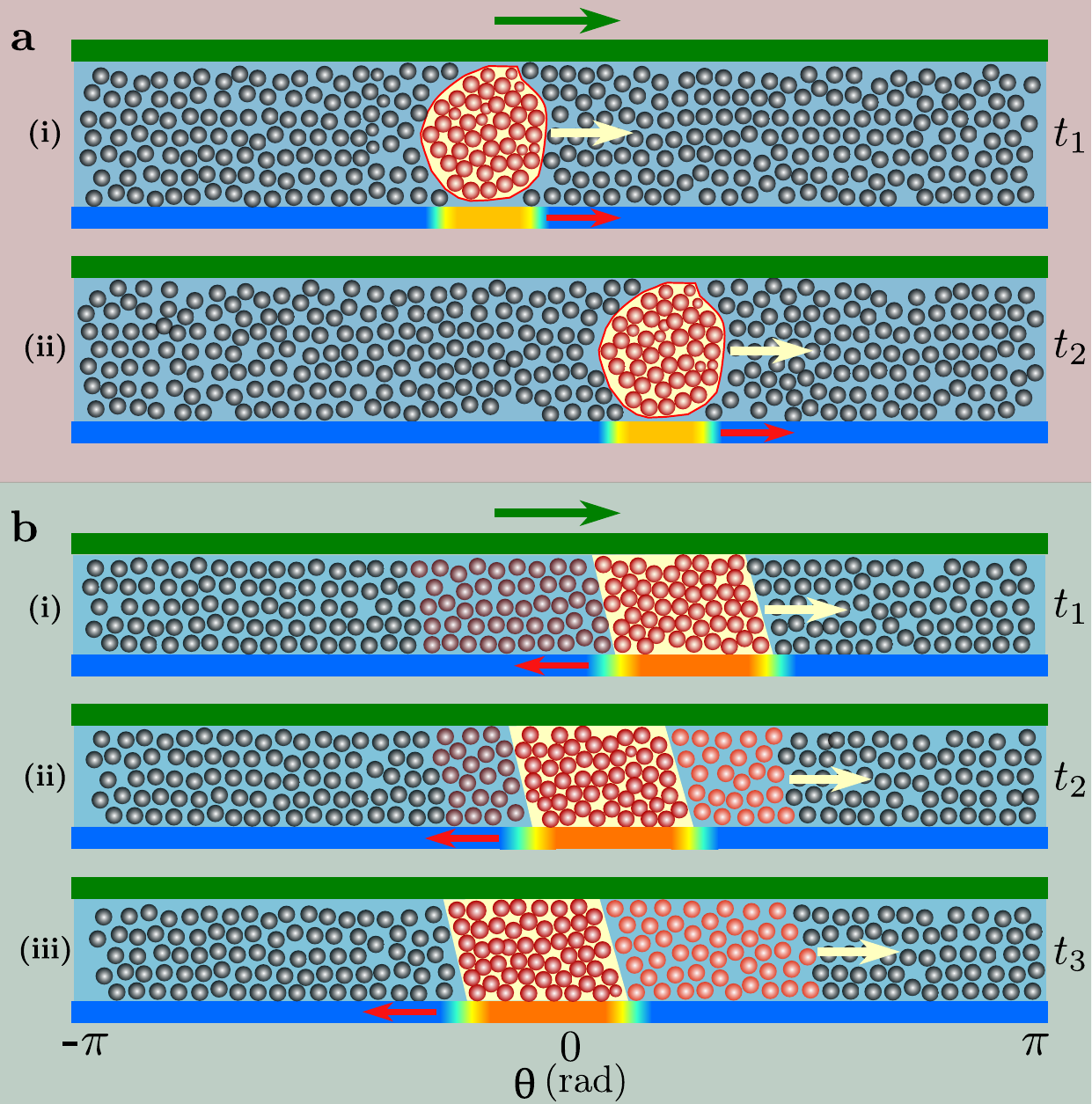}	
	\caption{{\bf Schematic illustrations of propagating high-stress domains in confined suspensions.}  {\bf (a)} A particle aggregation (red spheres) travels along the shear direction. If the aggregation remains anchored to the top plate, the particles travel together at the same speed (e.g., the case of PS-water suspensions at $h=1.10$~mm).  If the aggregation travels along the centerline of the gap, its propagation speed is approximately half that of the shear plate (e.g., the case of glass-oil suspensions). {\bf (b)} Under highly confined conditions, a particle aggregation may be partially jammed by the gap. As the top plate continues to shear,  the particles at the leading front move forward, and consequently disengage from the aggregation. Meanwhile, the particles trailing behind join the aggregation. This dynamic release-accumulation process leads to a counter-flow propagation of high-stress regions (e.g., the case of PS-water suspensions at $h=0.31$~mm). }
	\label{fig:model}
\end{figure}

The soft layers in our BSM setup were made of polydimethylsiloxane gels (PDMS from DMS-V31, Gelest Inc.) crosslinked by trimethylsiloxane copolymers (HMS-301, Gelest Inc.), with a Poisson's ratio $\nu=0.48$~\cite{Zhao2022}. The gel thickness was determined by the spin-coating speed applied to uncrosslinked PDMS mixtures, and the Young's modulus of a cured gel was controlled by the weight ratio of crosslinkers ($k$). We deposited a layer of 5~$\mu$m fluorescent beads on the PDMS surfaces as tracers to measure $u_x^*$ and $u_y^*$ and to quantify the boundary stresses $\sigma_{xz}$ and $\sigma_{yz}$ using Eq.~\ref{eq:tensor_product}. The objective was placed at a distance of $2R/3$ from the plate center, where the local shear rate was identical to the average shear rate measured by the rheometer (Eq.~\ref{eqn:rate}). The imaging speed was controlled between 3.8 frames per second and 7.6 frames per second, which was high enough to capture the evolving stress heterogeneity. As $\sigma_{xz}$ remained substantially larger than $\sigma_{yz}$ in all our experiments (Fig.~\ref{fig:BSM}(b) and Supplementary Figs.~S6 and S7), we  focused only on the evolution of $\sigma_{xz}$ in this study.

To characterize the PS-water suspensions, we chose $k = 0.97~\%$, $E = 7.5$~kPa, and a gel thickness $\delta_0=80$~$\mu$m. During the shear thickening transition, the average shear stresses ($\sim 10^1$~Pa) were large enough to remove the nanobeads from gel surfaces. To prevent this, we added an additional PDMS film with a thickness of 6~$\mu$m on the top (Fig.~\ref{fig:BSM}(a)), such that the gel surface ($\delta$) was slightly higher than the plane of nanobeads ($\delta_0$). We used $\sigma_{iz}$~$(i=x, y)$ at $z=\delta_0$ to estimate the boundary stress induced by the PS-water suspensions.

In contrast, the shear stresses in glass-oil suspensions were only $\sim 10^0$~Pa, substantially lower than those of PS-water suspensions. We thus chose $k = 0.81\%$ and no additional PDMS film was needed above the nano-beads ($\delta = \delta_0 = 80$~$\mu$m). To prevent the swelling effects due to the oil component in solvents, we swelled the gel layers with a 20~cSt silicone oil before the experiments. Consequently, the fully swollen PDMS layer has a thickness of 92~$\mu$m and a Young's modulus of $E= 1.7$~kPa.

{\em Ps-water suspensions.---} In the BSM setup, the boundary roughness decreased due to the presence of a soft layer on the bottom plate. Consequently, although size-dependent flow curves of PS-water suspensions still emerged at $h \leq 1.10$~mm, there were no non-monotonic behaviors even at $h=0.31$~mm, as shown in Fig.~\ref{fig:PS oscillations}(a). This result is consistent with our aforementioned observation that the rheology of PS-water suspensions is sensitive to boundary roughness. As the size-dependent rheology is the precursor of non-monotonic flow curves (Fig.~\ref{fig:S_curves}), we conjecture that the boundary stresses characterized in this regime ($h \leq 1.10$~mm) remain important indicators of the local flows in non-monotonic regimes.

By maintaining a constant shear stress  $\tau_a=20$~Pa in the thickening regime, we performed BSM measurements to characterize the evolution of local flows at $h=0.31$~mm, 1.10~mm, and 1.94~mm. Figure~\ref{fig:PS oscillations}(c) shows that $\langle \sigma_{xz}(t) \rangle$ remains steady at $h=1.94$~mm, but significantly fluctuates between a dominating low-stress state and a series of high-stress peaks under both $h = 1.10$~mm and $0.31$~mm.  The peak stresses of $\langle \sigma_{xz}(t) \rangle$ at $h=0.31$~mm are close to $200$~Pa, approximately twice those at $h=1.10$~mm, suggesting that the size-dependent flow curves were  associated with enhanced flow heterogeneity.

We further characterized the temporal intervals between adjacent high-stress spikes of $\langle \sigma_{xz}(t) \rangle$ at $h=1.10$~mm and $0.31$~mm. To quantify the characteristic timescale ($t_c$), we consider the autocorrelation function 
\begin{align}
 \notag R_\sigma(\delta t) & = \frac{ \mathrm{\bf E} [ \langle \sigma_{xz}(t) \rangle \langle \sigma_{xz}(t+\delta t) \rangle]}{\mathrm{\bf E}[\langle \sigma_{xz}(t) \rangle^2] } \\
			    & = \frac{\int_0^T \langle \sigma_{xz}(t) \rangle \langle \sigma_{xz}(t+\delta t) \rangle dt}{\int_0^T [\langle \sigma_{xz}(t) \rangle]^2 dt}
\label{eqn:autocorr}			    
\end{align}
where $T$ is the total measurement period. When $\delta t=t_c$, we expect a local maximum in the plot of $R_\sigma(\delta t)$. Figure~\ref{fig:PS oscillations}(c) shows the plots of $R_\sigma(\delta t)$ with the primary peaks at $t_{c1}=11.3$~s and $t_{c2}=13.0$~s for $h=1.10$~mm and 0.31~mm. These two timescales well captures the time intervals between high-stress peaks presented in Fig.~\ref{fig:PS oscillations}(b). Given the rotational period of the shear plate  $T_g = 4\pi R/ (3\dot\gamma h)$, we found 
\begin{equation}
 t_{c1} \approx T_g \:\:, \:\:\:\:\:\:\: t_{c2} \approx T_g/10.
\label{eqn:interval_PS}
\end{equation}
The relationships between $t_{c1(2)}$ and $T_g$ in Eq.~\ref{eqn:interval_PS} are important for understanding the underlying mechanisms of flow heterogeneity due to confinement. At $h=1.10$~mm, the approximate equality between $t_{c1}$ and $T_g$ aligns well with previous findings for the discontinuous shear-thickening of cornstarch-water suspensions~\cite{Rathee2020, Rathee2022}. However, the fast timescale $t_{c2}$ observed under the highly confined condition, $h=0.31$~mm, has not been reported.

\begin{figure*}[t]
	\centering
	\includegraphics[width = 160 mm]{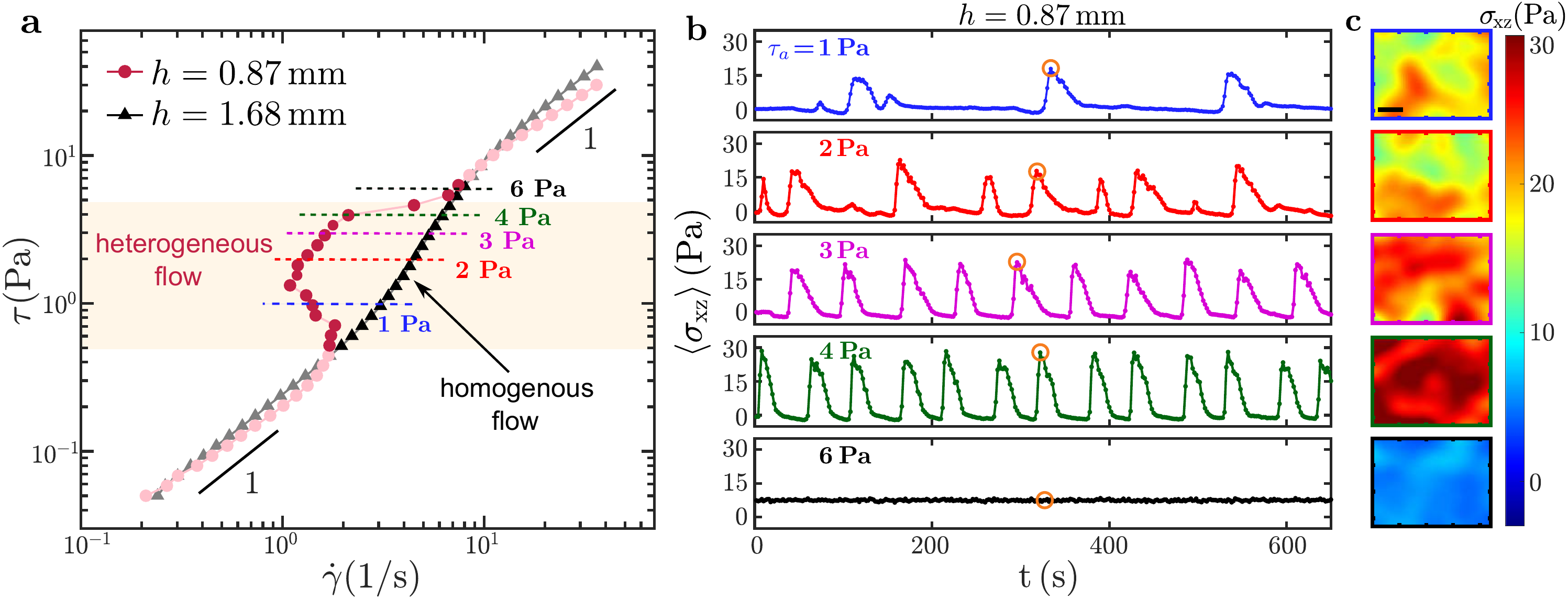}	
     \caption{{\bf Local stress fluctuations in  glass-oil suspensions}. {\bf (a)} Flow curves measured with the BSM setup for $h = $ 1.68 and 0.87~mm, respectively. The dashed lines indicate the shear stresses ($\tau_a$) that we applied to suspensions while conducting long-period BSM measurements. {\bf (b)} Temporal fluctuations of $\langle \sigma_{xz} (t) \rangle$ at $\tau_a = 1$~Pa, 2~Pa, 3~Pa, 4~Pa, and 6 Pa for  $h = $ 0.87~mm. {\bf (c)} Local stress maps $(\sigma_{xz})$ at the times indicated by the open circles in (b). Scale bars: $200$~$\mu$m.}
	\label{fig:glass-oil oscillations}
\end{figure*}

\begin{figure}[h]
	\centering
	\includegraphics[width = 75 mm]{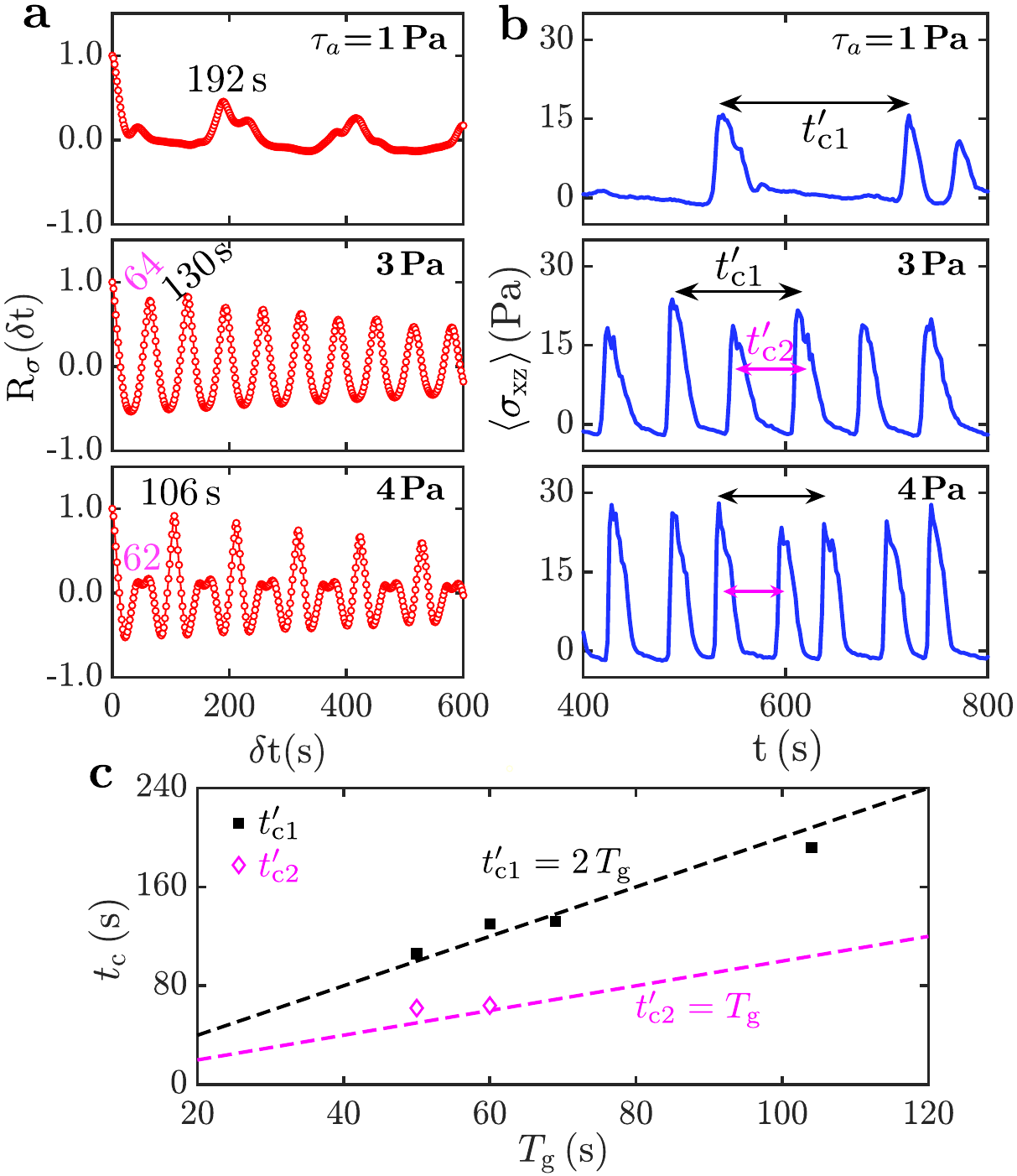}	
	\caption{{\bf Fluctuation timescales of local shear stress ($\sigma_{xz}$) in confined glass-oil suspensions.} {\bf (a)} Autocorrelation functions $R_\sigma(\delta t)$ at $\tau_a = 1$~Pa, 3~Pa, and 4~Pa for $h=0.87$~mm. {\bf (b)} Plots of $\langle \sigma_{xz}(t) \rangle $ at $\tau_a = 1$~Pa, 3~Pa, and 4~Pa. The two characteristic timescales, $t_{c1}^\prime$ and $t_{c2}^\prime$, are indicated by the black and pink arrows, respectively. {\bf (c)} Plots of $t_{c1}^\prime$ and $t_{c2}^\prime$ against the rotational period of the shear plate ($T_g$).}
\label{fig:timescale} 
\end{figure}

To address the difference between $t_{c1}$ and $t_{c2}$, we investigated the propagations of local high-stress domains. Figure~\ref{fig:PS front} shows the snapshots of  $\sigma_{\rm xz}$ evolving during representative high-stress events for the two gap sizes. At $h=1.10$~mm, a millimeter-sized high-stress front emerges from the left, and then moves rightward to cross the imaging field (Fig.~\ref{fig:PS front}(a) and Supplementary Video 1), aligning with the shear direction.  In contrast, we observed reversed propagations of high-stress regions at $h=0.31$~mm, with these regions initially appearing from the right and then traveling to the left, as shown in Fig~\ref{fig:PS front}(b) for two typical events (Supplementary Video S2).  This counter-shear propagation with a traveling period ($t_{c2}\approx T_g/10$) was common for $h<0.5$~mm. Compared with previous observations in cornstarch-water suspensions~\cite{Hermes2016,Rathee2020},  the counter-flow of high-stress regions was more stable in confined PS-water suspensions.

As illustrated in Fig.~\ref{fig:model}, we propose two flow mechanisms to interpret the propagations of high-stress domains in different directions. For the PS-water suspensions at $h = 1.10$~mm, we conjecture that a locally jammed, solid-like particle aggregation moves along the shear direction (Fig.~\ref{fig:model}(a)). Under no-slip boundary conditions, the propagation speed of this particle aggregation remains close to that of the shear plate. Consequently,  a high-stress domain emerges once every rotational period ($t_{c1} \approx T_g$)~\cite{Rathee2022}. In contrast, we speculate that the counter-flow propagation at $h=0.31$~mm is associated with a simultaneous accumulation-release process of particle aggregations. As illustrated by Fig.~\ref{fig:model}(b), the particle aggregations can be partially jammed by confinement at $h=0.31$~mm. While the particles at the leading front migrate forward and gradually separate from the aggregation (yellow arrows in Fig.~\ref{fig:model}(b)), the trailing particles are stopped by this locally jammed region (red arrows in Fig.~\ref{fig:model}(b)), which results in a  propagation of high-stress regions against the shear direction. As this backward propagation of boundary stresses is determined by the particle-density, the propagation speed can be significantly higher than that of the shear plate ($t_{c2} \approx T_g/10$). In both scenarios (Figs.~\ref{fig:model}(a) and (b)), we expect enhanced local dilations of the high-stress domains~\cite{Hermes2016, Rathee2022}.

{\em Glass-oil suspensions.---} We further performed BSM measurements on glass-oil suspensions. Since the rheological behaviors of glass-oil suspensions were insensitive to the boundary roughness (Sec.~III and Supplementary Fig.~S5(b)), the non-monontonic flow curves remained unchanged in the BSM setup (Fig.~\ref{fig:glass-oil oscillations}(a)). The material properties of glass-oil suspensions have two advantages for BSM characterization: (1) the non-volatile solvent allows {\em in-situ} measurements over a long period, and (2) the glass-oil suspensions exhibits a broader shear-thickening range than the PS-water suspensions. Thus, for glass-oil suspensions, we were able to conduct BSM measurements by systematically varying the shear stress within shear thickening regime. 

Figure~\ref{fig:glass-oil oscillations}(a) shows that the non-monotonic flows of glass-oil suspensions emerges as the gap size decreases from $h=1.68$~mm to $h=0.87$~mm in the BSM setup. We maintained the global shear stress constant at $\tau_a = 1$~Pa, 2~Pa, 3~Pa, 4~Pa, and 6~Pa, respectively. Figure~\ref{fig:glass-oil oscillations}(b) presents the temporal evolutions of the averaged local stress $\langle \sigma_{xz} \rangle$ under the confined condition, $h=0.87$~mm. Similar to PS-water suspensions, the local stress $\langle \sigma_{xz} \rangle$ in glass-oil suspensions also fluctuates between a nearly zero-stress state and a series of high-stress peaks within the non-monotonic regime. As $\tau_a$ increases from 1~Pa to 4~Pa, the high-stress peaks gradually rise from 15~Pa to 30~Pa. Representative stress maps are shown in Fig.~\ref{fig:glass-oil oscillations}(c). However, as $\tau_a$ further increases to  6~Pa, the flow heterogeneity notably reduces and $\sigma_{xz}$ remains constant around 7.8~Pa (Supplementary Video S4). We interpret this homogenization of shear flows at $\tau_a = 6$~Pa as a dynamic yielding process in that the high-stress domains relax under shear. Consequently, this process substantially reduces flow resistance (Fig.~\ref{fig:glass-oil oscillations}(a)) and  decreases normal stresses sharply (Fig.~\ref{fig:S_curves}(e)).

At a large gap size $h=1.68$~mm, the local stress fluctuations remain insignificant throughout the shear thickening regime, and the BSM measurements show a nearly homogenous shear flow (Supplementary Fig.~S7). These findings further confirm that the flow heterogeneity depicted in Fig.~\ref{fig:glass-oil oscillations} was induced by boundary confinement.

\begin{figure*}[t]
	\centering
	\includegraphics[width = 160 mm]{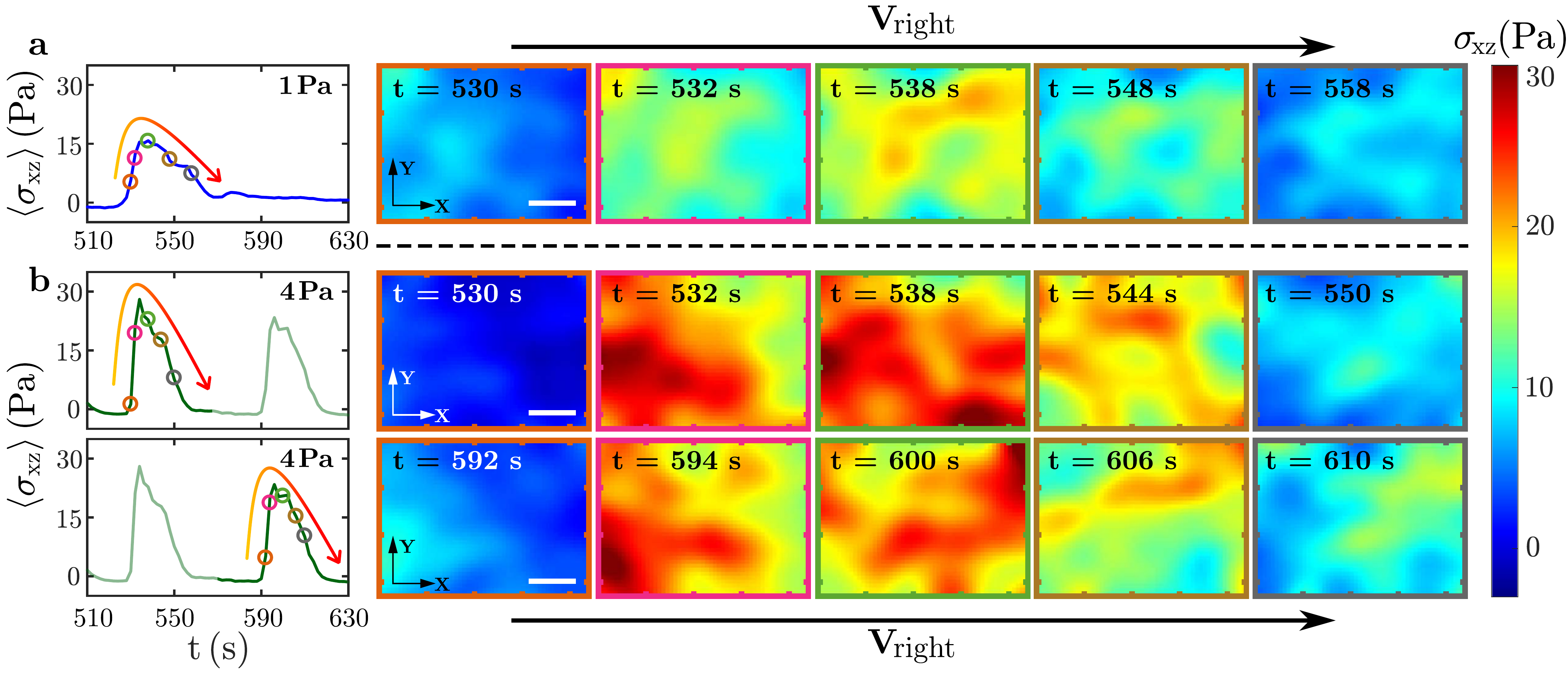}	
	\caption{{\bf Propagation of  high-stress regions in confined glass-oil suspensions at {\bf (a)} $\tau_a = 1$~Pa and {\bf (b)} $\tau_a = 4$~Pa}. Left panels: representative peaks of $\langle \sigma_{xz} (t) \rangle$ at $h=0.87$~mm. Right panels:  stress maps of $\sigma_{xz}$ at different times indicated by the open circles in the left panels. Scale bars: $200$~$\mu$m.} 
	\label{fig:glass_front}
\end{figure*}

Using the autocorrelation analysis employed in Eq.~\ref{eqn:autocorr}, we quantified the characteristic timescales between two consecutive high-stress peaks for confined glass-oil suspensions ($h=0.87$~mm).  Figure~\ref{fig:timescale}(a) represents the results of $R_\sigma$ evaluated at $\tau_a=1$~Pa, 3~Pa, and 4~Pa. When $\tau_a = 1$~Pa, the first peak in $R_\sigma(\delta t)$ is located at $\delta t= 192$~s, which well approximates the time interval between two adjacent peaks ($t_{c1}^\prime$) in Fig.~\ref{fig:timescale}(b). For $\tau_a =3$~Pa and 4~Pa, $\langle \sigma_{xz}(t) \rangle$ oscillates more rapidly with two distinct timescales. For example, we observed a local maximum at $\delta t = 62$~s followed by a major peak at $\delta t =104$~s. As shown in Fig.~\ref{fig:timescale}(b), the longer timescale corresponds to the intervals between two major peaks ($t_{c1}^\prime \approx 106$~s), whereas the shorter timescale approximates a sub-interval between a lower spike and a major peak ($t_{c2}^\prime \approx 62$~s). The two intervals are indicated by the black and pink arrows.

We further compared both $t_{c1}^\prime$ and $t_{c2}^\prime$ with the rotational period $T_g$. Figure~\ref{fig:timescale}(c) shows the plots of both $t_{c1}^\prime$ and $t_{c2}^\prime$ against $T_g$. For $\tau_a=1$~Pa and 2~Pa, only the timescale of $t_{c1}^\prime$ was identified in the autocorrelation analysis. In contrast, for $\tau_a=3$~Pa and 4~Pa, both $t_{c1}^\prime$ and $t_{c2}^\prime$ appear in the plots of $R_\sigma(\delta t)$. The results in Fig.~\ref{fig:timescale}(c) indicate two linear scalings, 
\begin{equation}
 t_{c1}^\prime \approx 2 T_g \:\:, \:\:\:\:\:\:\: t_{c2}^\prime \approx T_g. 
\label{eqn:interval_glass}
\end{equation}
In contrast with the characteristic timescales measured for the PS-water suspensions (Eq.~\ref{eqn:interval_PS}), the ultra-fast propagating mode ($\sim T_g/10$) was not observed in the glass-oil suspensions (Eq.~\ref{eqn:interval_glass}). 

To interpret the timescales $t_{c1}^\prime$ and $t_{c2}^\prime$, we also measured the dynamics of the high-stress domains in the confined glass-oil suspensions. Figure~\ref{fig:glass_front} shows the propagation of representative high-stress domains for $\tau_a=1$~Pa and 4~Pa, respectively. These local domains always travel in the same direction as the shear plate, and no counter-flow propagation was observed in any experiments (Supplementary Videos S5). We conjecture that the simultaneous accumulation and release of high-stress domains (as illustrated in Fig.~\ref{fig:model}(b)) are absent in glass-oil suspensions due to the relatively weak stress localizations ($\langle \sigma_{xz}\rangle < 30$~Pa) in the non-monotonic regime. 

We herein interpret the dependence of $t_{c1}^\prime$ and $t_{c2}^\prime$ on $T_g$ (Eq.~\ref{eqn:interval_glass}) based on the flow mechanisms shown in Fig.~\ref{fig:model}(a). When $\tau_a = 1$~Pa or 2~Pa, briefly exceeding the onset of shear thickening, a local particle aggregation forms in the suspensions. Due to the weak boundary stresses ($\langle \sigma_{xz} \rangle < 30$~Pa),  the particle aggregation possibly travels along the center of the shear cell without sticking to the shear plate~\cite{Guillaume2020}. The mean traveling speed of the aggregation is approximately half that of the shear plate, such that he high-stress region finishes one complete round during two rotational periods, $t_{c1}^\prime \approx 2 T_g$. For $\tau_a =3$~Pa or 4~Pa, a second  high-stress domain can appear in a different location within the suspension. We thus observed the high-stress regions twice as frequently as we did at lower stresses, leading to $t_{c2}^\prime \approx T_g$.

\section{Conclusions}

By systematically characterizing the shear thickening behaviors of both non-sedimenting PS-water suspensions and sedimenting glass-oil suspensions, we have demonstrated the critical role of boundary confinement in determining shear thickening rheology. The major findings of this work are summarized as follows.

(1)~Although non-monotonic flow curves have been predicted theoretically, they have rarely been reported in previous experiments due to flow instabilities~\cite{Han2019, Olmsted2008}. Our results show that strong confinement imposed by a shear boundary can effectively induce stable $S$-shaped responses in the shear thickening regime (Fig.~\ref{fig:S_curves}). Although  high-stress domains develop in the non-monotonic regimes,  boundary confinement prevents the stress heterogeneity from evolving into large-scale rheochaos~\cite{Hermes2016}.

(2)~Using boundary stress microscopy (BSM) (Fig.~\ref{fig:BSM}), we characterized the rich dynamics of local stress heterogeneity induced by boundary confinement. For both  PS-water and glass-oil suspensions, we observed the propagations of high-stress domains at speeds controlled by the rotational shear plate (Eqs.~\ref{eqn:interval_PS} and~\ref{eqn:interval_glass}). For PS-water suspensions, we observed a crossover from the shear-directed to counter-shear propagations of high-stress domains as the gap size reduced (Fig.~\ref{fig:PS front}). While both propagation directions have previously been reported in dense suspensions~\cite{Guillaume2020, Rathee2017,Rathee2020,Gauthier2021,Hermes2016}, we demonstrated that these local high-stress domains remain stable under confinement. Intriguingly, the counter-shear propagation in PS-water suspensions appears to be significantly faster than the rotation of the shear plate (Eq.~\ref{eqn:interval_PS}), potentially caused by a simultaneous accumulation-release process of high-stress aggregations (Fig.~\ref{fig:model}(b)). For glass-oil suspensions, however, the high-stress domains always propagate along the shear direction, regardless of the gap size (Fig.~\ref{fig:glass_front}). The absence of counter-shear propagation is possibly attributed to the weak boundary stresses ($\sigma_{xz}\sim 30$~Pa) induced in confined glass-oil suspensions .

(3) The rheology of PS-water suspensions depends on the material properties of shear boundaries. With shear plates made of roughened glass, $S$-shaped flow curves appeared for confined PS-water suspensions (Fig.~\ref{fig:S_curves}(b)). However, when a gel film was coated on the bottom plate for BSM measurements, the flow curves were consistently monotonic even under strong confinement (Fig.~\ref{fig:PS oscillations}(a)). We attribute this absence of non-monotonic responses to the smoothness of gel surfaces and the deformations of gel films under large boundary stresses ($\sigma_{xz}\sim 300$~Pa) in confined PS-water suspensions. 

In summary,  we presented an experimental investigation that revealed the underlying connection between shear thickening rheology and local flow structures in confined dense suspensions. We have demonstrated that boundary confinements can effectively stabilize $S$-shaped flow curves and induce local stress heterogeneity. Our results provide valuable insights into controlling suspension rheology through boundary effects.

\section*{Acknowledgement}
We thank Prof.~Ryohei Seto for valuable suggestions. This work was supported by the Early Research Scheme (26309620), General Research Funds (No.16307422 and No.16305821),  and the Collaborative Research Fund (No.CY6004-22Y) from the Hong Kong Research Grants Council. The research activities were also funded by the Partnership Seed Fund (No.ASPIRE2021$\#$1) from the Asian Science and Technology Pioneering Institutes of Research and Education League and Hong Kong-Macau-Guangdong  Industrialization Fund  from Guangdong Science and Technology Department (No.2023A0505030017). 
\bibliography{reference}

\end{document}